# Voltage control of magnon spin currents in antiferromagnetic $Cr_2O_3$

Changjiang Liu[1†], Yongming Luo[1,2†], Deshun Hong[1], Steven S.-L. Zhang[1,3], Brandon Fisher[4], John E. Pearson[1], J. Samuel Jiang[1], Axel Hoffmann[1,5*] and Anand Bhattacharya[1*]

[1] *Materials Science Division, Argonne National Laboratory, Argonne, Illinois 60439, USA,*
[2] *School of Electronics and Information, Hangzhou Dianzi University, Hangzhou, Zhejiang, 310018, China,*
[3] *Department of Physics, Case Western Reserve University, Cleveland, Ohio 44106, USA,*
[4] *Nanoscale Science and Technology Division, Argonne National Laboratory, Lemont, IL 60439, USA,*
[5] *Department of Materials Science and Engineering, University of Illinois at Urbana-Champaign, Urbana, Illinois 61801, USA.*

*\* Correspondence to anand@anl.gov or axelh@illinois.edu*
[†] *Denotes equal contribution*

*Abstract*

Voltage-controlled spintronic devices utilizing the spin degree of freedom are desirable for future applications, and may allow energy-efficient information processing. Pure spin current can be created by thermal excitations in magnetic systems via the spin Seebeck effect (SSE). However, controlling such spin currents, only by electrical means, has been a fundamental challenge. Here, we investigate the voltage control of the SSE in the antiferromagnetic insulator $Cr_2O_3$. We demonstrate that the SSE response generated in this material can be effectively controlled by applying a bias voltage, owing to the sensitivity of the SSE to the orientation of the magnetic sublattices as well as the existence of magnetoelectric couplings in $Cr_2O_3$. Our experimental results are explained using a model based on the magnetoelectric effect in $Cr_2O_3$.

In addition to electric charge, the spin degree of freedom of electrons can be used to carry and process information. Because transport of spins may be realized in the form of magnon currents without movement of charges, spintronics has the potential to provide the next generation of computing with low power dissipation [1,2]. With extensive research efforts made over the last few years, a variety of methods for generating spin currents have been developed [2-5]. One of them is to use thermal excitations of magnons inside magnetic insulators, which is also known as the spin Seebeck effect (SSE) [6]. Since its discovery, the SSE has found a versatile role in generating pure spin currents in diverse material systems, including ferrimagnetic [7], paramagnetic [8,9], and antiferromagnetic materials [10,11]. This is because the SSE does not require long-range magnetic order and coherent precession at resonance, both of which are necessary for other methods such as spin pumping [9,12-14]. Because of this unique property, the SSE is convenient for converting heat, an inevitable byproduct during information processing, into useful spin currents. To further incorporate the SSE into spintronic applications, one also desires effective ways to electrically control the SSE signal, an area that is still largely unexplored. Antiferromagnetic insulators are promising platforms for this purpose since in addition to their low stray fields and potentially high operation speeds [15,16], some antiferromagnets have magnetoelectric or multiferroic properties, which allow for direct coupling between the spin lattices and electric fields [17,18]. Thus, these materials are particularly attractive as platforms for controlling magnon spin currents with electric fields [19,20].

In this letter, we investigate the electrical control of the SSE in the antiferromagnetic insulator $Cr_2O_3$, which is one of the first antiferromagnets in which the SSE was demonstrated [10]. The spins of $Cr^{3+}$ ions in neighboring layers align antiferromagnetically with the [0001] axis; the antiferromagnetic order breaks time-reversal symmetry, while the staggered crystal field for $Cr^{3+}$ ions breaks space-inversion symmetry [21]. Application of an electric field along the [0001] axis displaces the $Cr^{3+}$ ions relative to the ligand $O^{2-}$, which breaks the equivalence of the two magnetic sublattices of $Cr_2O_3$. As a result, a finite magnetization can be induced in $Cr_2O_3$ with an electric field due to a net ferrimagnetic spin arrangement. This phenomenon is known as the magnetoelectric (ME) effect, which was first discovered in the 1960s [22,23]. Because of the coupling between the spin-lattice and external electric fields, much effort has been devoted to exploring possible applications of $Cr_2O_3$ in voltage-controlled spintronics devices [24-26]. In recent years, by using the exchange bias and/or anomalous Hall effects as read-out techniques [27-33], it has been shown that electrical switching of magnetic sublattices in thin-film $Cr_2O_3$ is possible. Since in SSE measurements, the quantization axis of the magnon spin current is naturally determined by the orientation of the magnetic lattices [12,34], the electrical manipulation of these in $Cr_2O_3$ would allow for control of the SSE response.

To accomplish electrical manipulation of the SSE, devices that incorporate $Cr_2O_3$ films with both top and bottom electrodes were fabricated on (0001) orientated $Al_2O_3$ substrate by dc

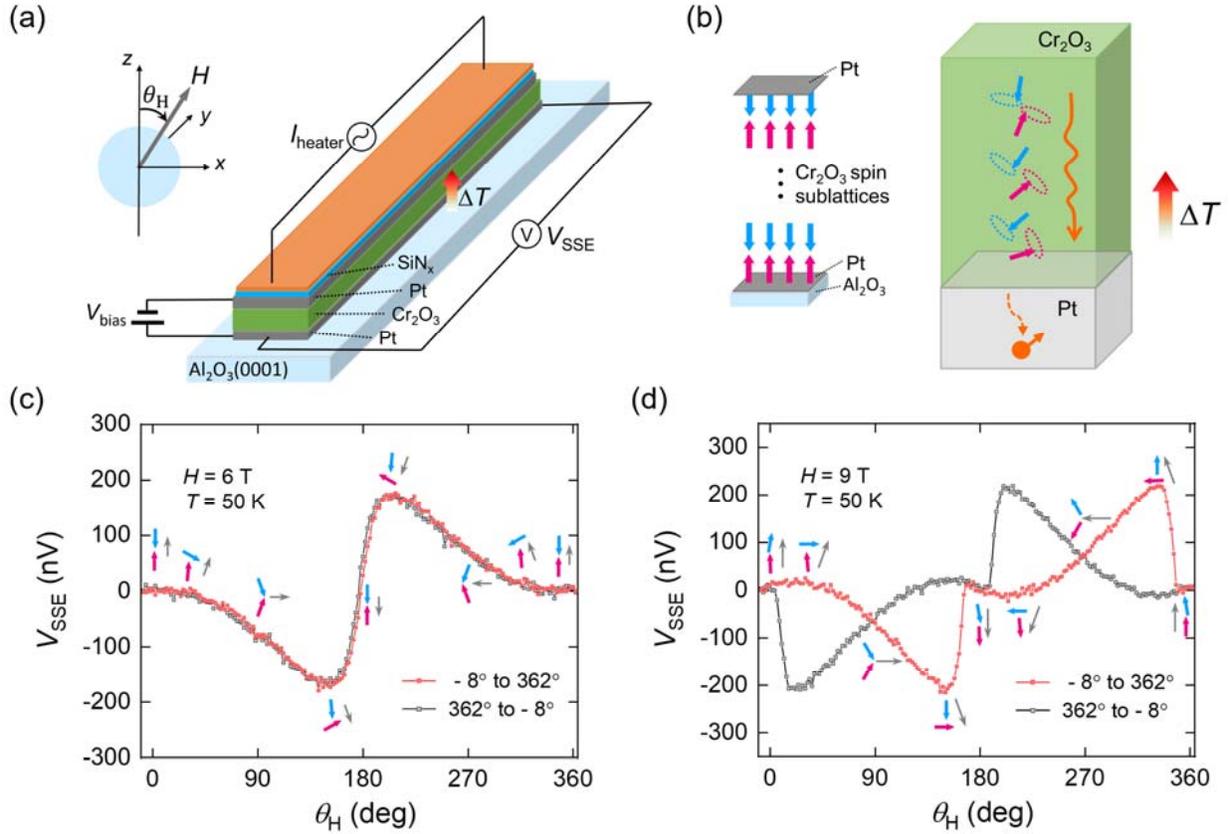

Figure 1. Device design, measurement geometry, and SSE data. (a) Schematics of the SSE device showing the vertical structure of the sample. The direction of the magnetic field is varied in the $x-z$ plane and the spin Seebeck voltage $V_{SSE}$ develops in the $y$-direction. (b) Left panel: at zero magnetic field, the two magnetic sublattices of $Cr_2O_3$ are aligned along the $z$-direction. The sublattice in direct contact with the bottom Pt layer is shown in red. Right panel: magnon spin currents created by thermal excitations in $Cr_2O_3$ (under a tilted magnetic field) propagates towards the $Cr_2O_3$/Pt interface (orange wiggly arrow). A spin current (dashed arrow) carried by conduction electrons in the Pt layer is converted from the magnon current in the $Cr_2O_3$ layer at the interface. (c) and (d) SSE signal measured as a function of the angle of the magnetic field, for fields below and above the surface spin-flop field, respectively. The red and blue arrows represent the orientations of the bottom and top surface spins, respectively. They correspond to the red trace of the SSE data in (d) (clockwise field rotation). Gray arrows indicate the orientations of the magnetic field.

and rf reactive sputtering. Figure 1(a) shows the vertical structure of our SSE devices. A 250-nm thick $Cr_2O_3$ film is sandwiched between a bottom and top Pt layers, which allow for applying a bias voltage along the $c$-axis of $Cr_2O_3$. An on-chip heater is deposited on top, electrically isolated from the Pt/$Cr_2O_3$/Pt trilayer by a 100-nm thick film of $SiN_x$. During the measurement, a sinusoidal electrical current with a frequency of 3 Hz is applied through the heater wire, which generates a temperature gradient in the vertical direction as indicated by the red arrow in

Fig. 1(a). Magnons excited inside $Cr_2O_3$ propagate towards/away from the two $Cr_2O_3$/Pt interfaces when a temperature gradient develops [Fig. 1(b)]. The spin current is then measured as a transverse voltage $V_{SSE}$ by a lock-in amplifier at the second harmonic in the bottom Pt layer via the inverse spin Hall effect (ISHE).

In zero magnetic field, the two antiferromagnetically coupled spin lattices in $Cr_2O_3$ are collinear with the *c*-axis, which is equivalent to the *z*-direction of our measurement geometry [see Figs. 1(a) and (b)]. An applied magnetic field breaks their equivalence and when applied in a direction other than the *c*-axis results in tilting the orientations of both magnetic sublattices. During SSE measurements, the direction of the external magnetic field, indicated by $\theta_H$, is rotated in the $x - z$ plane relative to the sample. The measured $V_{SSE}$ displays unique angular dependencies, which are shown in Figs 1(c) and (d) for $H = 6$ T and 9 T, respectively. Data corresponding to $\theta_H$ rotated in the clockwise ($-8°$ to $362°$) and counterclockwise ($362°$ to $-8°$) directions are shown in red and black, respectively. Distinct differences in the angular dependences of $V_{SSE}$ are observed between Figs. 1(c) and (d): at $H = 6$ T, $V_{SSE}(\theta_H)$ is reversible (i.e., red and black traces overlap), while at $H = 9$ T, $V_{SSE}(\theta_H)$ displays clear hysteretic behavior.

The hysteretic $V_{SSE}(\theta_H)$ seen in Fig 1(d) is due to the switching of the two magnetic sublattices in $Cr_2O_3$ over the magnetic hard plane, together with a slight imbalance between the two sublattices (Supplemental Note 1), in response to the rotating magnetic field. This happens when the magnitude of the applied field is greater than the surface spin-flop field, which cants the two sublattices relative to each other, creating a finite magnetic moment that induces a net torque on the two sublattices. The two spin sublattices also experience a uniaxial magnetic anisotropy [36], with the *c*-axis being the easy axis. When the magnetic field is low, the net torque on the magnetic sublattices is unable to overcome the anisotropy field, and the orientations of the sublattice spins only tilt back and forth about the *c*-axis as the magnetic field is rotated, producing a non-hysteretic $V_{SSE}$ response as seen in Fig. 1(c).

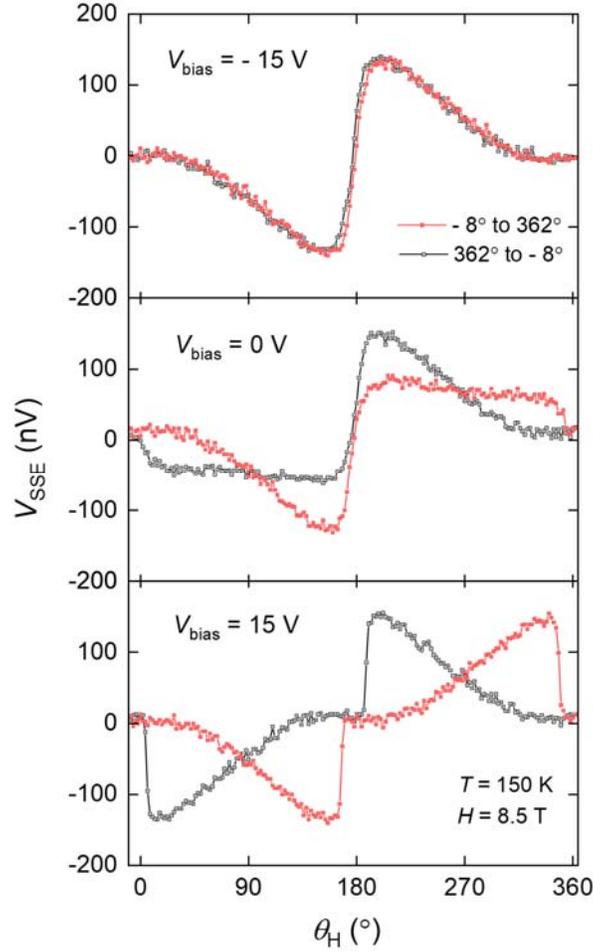

Figure 2. Control of the angular dependence of the SSE using a bias voltage. Angular dependence of the $V_{SSE}$ measured under bias voltages of – 15 V (top panel), 0 V (middle panel), and 15 V (bottom panel), respectively. The strength of the magnetic field is constant at $H = 8.5$ T in all measurements.

In recent work, we studied the spin orientations in more detail by taking into account the finite thickness of the $Cr_2O_3$ and the concomitant surface spin-flop transition [35]. Schematics of the orientations of the spins in the bottom (red) and top (blue) surface are illustrated in Figs. 1(c) and (d) along with the SSE data. Full spin structures throughout the thickness are presented in the Supplemental Materials. As seen in Fig. 1(d), for $H = 9$ T the spin orientations are reversed after hysteretic switching when comparing the spin configurations at $\theta_H = 0$ and 180 degrees, while they are identical at these two field angles for the 6-T measurement shown in Fig. 1(c). The angular dependence of the $V_{SSE}$ follows the horizontal component of the surface spins that are in direct contact with the bottom Pt layer, which is our ISHE detector. This is because spins on the (0001) surface of $Cr_2O_3$ are uncompensated, which act as a spin polarizer for the bulk magnon spin currents reaching the $Cr_2O_3$/Pt interface; and the geometry of our ISHE measurement is only sensitive to spins polarized along the $x$ – direction (transverse to $V_{SSE}$).

Next, we investigated the voltage-control of the SSE by applying a bias voltage along the $c$-axis of $Cr_2O_3$. A positive bias is defined such that the electric field points from the bottom to the top surface. Our measurements show a clear bias-voltage ($V_{bias}$) control of the $V_{SSE}$, which is presented in Fig. 2. At $V_{bias} = 0$ V (middle panel of Fig. 2), the $V_{SSE}$ shows partial switching behavior at $H = 8.5$ T, as indicated by the relatively weak hysteretic angular dependence. This might be due to the existence of domains in the $Cr_2O_3$ film [37] caused by spatial variations in anisotropy energies. Strikingly, when a finite bias voltage is applied, the angular response of $V_{SSE}$ is completely modified, as shown in the upper and lower panels for negative and positive bias voltages, respectively. The negative bias voltage fully suppresses the switching of the magnetic sublattices as indicated by the continuous $V_{SSE}$ in response to the rotation of $\theta_H$; while the positive bias results in a jump and full hysteretic curve in $V_{SSE}$.

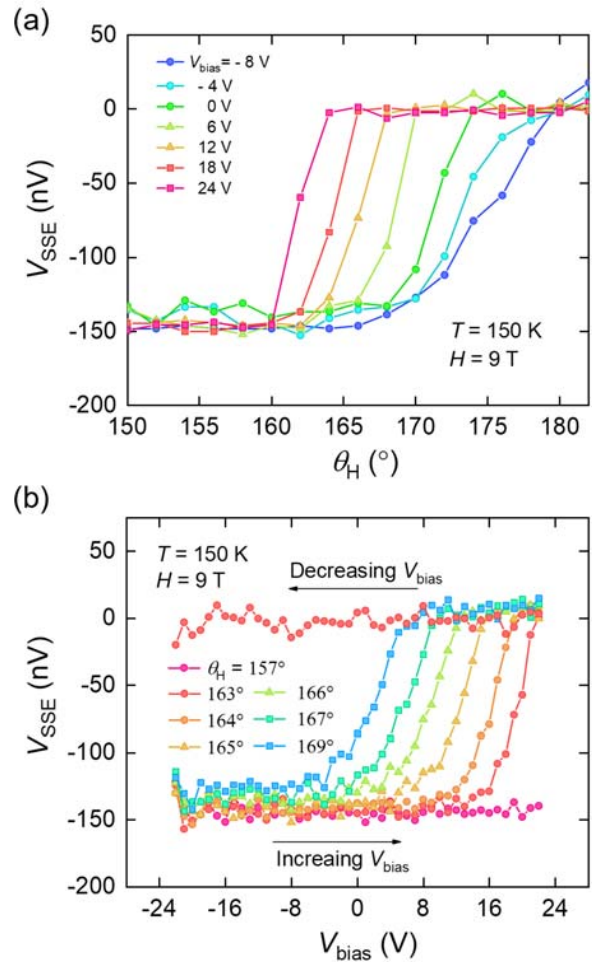

Figure 3. Bias-voltage control of $V_{SSE}$. (a) Angular dependence of $V_{SSE}$ measured near the spin-flop transition under different bias voltages. (b) Transition of $V_{SSE}$ driven by the bias voltage under a fixed magnetic field strength applied at along several different angles $\theta_H$. The sweep-direction of the bias voltage is indicated by the black arrows. With increasing $V_{bias}$, $V_{SSE}$ are driven to switch from about – 150 nV to 0 nV at different voltages depending on $\theta_H$.

To further explore the bias-voltage control of $V_{SSE}$, we have performed a detailed angular dependence measurement of $V_{SSE}$ near the surface spin-flop transition in a range of bias voltages, which are shown in Fig. 3(a). The critical magnetic field angle $\theta_c$ where $V_{SSE}$ displays a jump shows a strong dependence on the applied bias voltage. When $V_{bias}$ is positive, $\theta_c$ is pushed to lower values (easy to switch). For example, at $V_{bias} = 24$ V, $\theta_c$ is reduced to 162° as compared to $\theta_c = 171°$ at $V_{bias} = 0$ V. Here, $\theta_c$ is determined using the magnetic field angle where the total change in $V_{SSE}$ is half of its maximum. Under positive biases, the transition in $V_{SSE}$ also becomes sharper. In contrast, applying negative bias voltages produces the opposite effect, i.e., the transition in $V_{SSE}$ is hindered – with the $\theta_c$ pushed to higher values.

The wide tunability of $\theta_c$ by $V_{bias}$ suggests that one can drive the transition in $V_{SSE}$ by sweeping only $V_{bias}$ while keeping the magnetic field $H$ fixed. The measurement results are shown in Fig. 3(b). A magnetization configuration independent background voltage coming from leakage currents through the $Cr_2O_3$ film has been subtracted from the raw data (see Supplemental Fig. S3). During the measurement, the magnetic sublattices are first initialized by rotating the magnetic field (9 T) from 0° to an angle close to $\theta_c$ as determined from the angular dependence of $V_{SSE}$ shown in Fig. 3(a). $V_{bias}$ is then swept from negative to positive values, during which we observe a clear voltage-driven transition in $V_{SSE}$ arising from the reversal of magnetic sublattices. For instance, with $\theta_H = 163°$ as shown in Fig. 3(b), $V_{SSE}$ rises from about $-150$ nV to 0 nV as $V_{bias}$ increases from 15 V to 22 V. When $\theta_H$ is set to higher angles [closer to $\theta_c(V_{bias} = 0$ V$)$], the corresponding bias voltage required to drive the transition in $V_{SSE}$ becomes lower. They are shown in Fig. 3(b) in different colors. For the measurement at $\theta_H = 169°$, a negative $V_{bias}$ of $-20$ V was first applied to inhibit the transition during the initialization of magnetic field because $\theta_H$ is close to $\theta_c (V_{bias} = 0$ V$)$.

The electrical manipulation of the SSE presented above can be understood based on the ME coupling in $Cr_2O_3$. Figure 4(a) shows the lattice and spin structures of $Cr_2O_3$ in a primitive cell along the $c$-axis. Adjacent $Cr^{3+}$ ions are separated by either two ligand $O^{2-}$ triangles or just a single $O^{2-}$ triangle with a slightly smaller size. The application of an electric field along the $c$-axis will drive those $Cr^{3+}$ ions (shown with red spins) closer to the double $O^{2-}$ triangle and away from the single one, while the movement of the other $Cr^{3+}$ ions (shown with blue spins) relative to their ligand $O^{2-}$ triangles are opposite [38]. These changes in bond lengths between the $Cr^{3+}$ and ligand $O^{2-}$ ions break the equivalence of the two magnetic sublattices, which induces a net magnetization $\Delta m = \alpha_{zz} E_z$, where $\alpha_{zz}$ is the diagonal component of the linear ME coefficient, and $E_z$ is z-component of the electric field [39-41]. To illustrate the process of the switching of the sublattices more clearly, we propose that under a positive (negative) electric field, the absolute magnitude of the magnetic moment of the sublattice A, shown with red arrows in Fig. 4(a), would increase (decrease), while the magnetic moment of the sublattice B would decrease (increase), regardless of the orientation of their spins. In other words, the electric field gives rise to an effective ferrimagnetic spin configuration. Note that, under this scenario, the sign of $\alpha_{zz}$ is dependent on the orientations of the two sublattices, which is consistent with experimental findings, i.e., configuration of Fig. 4(a) corresponds to a positive $\alpha_{zz}$ [42].

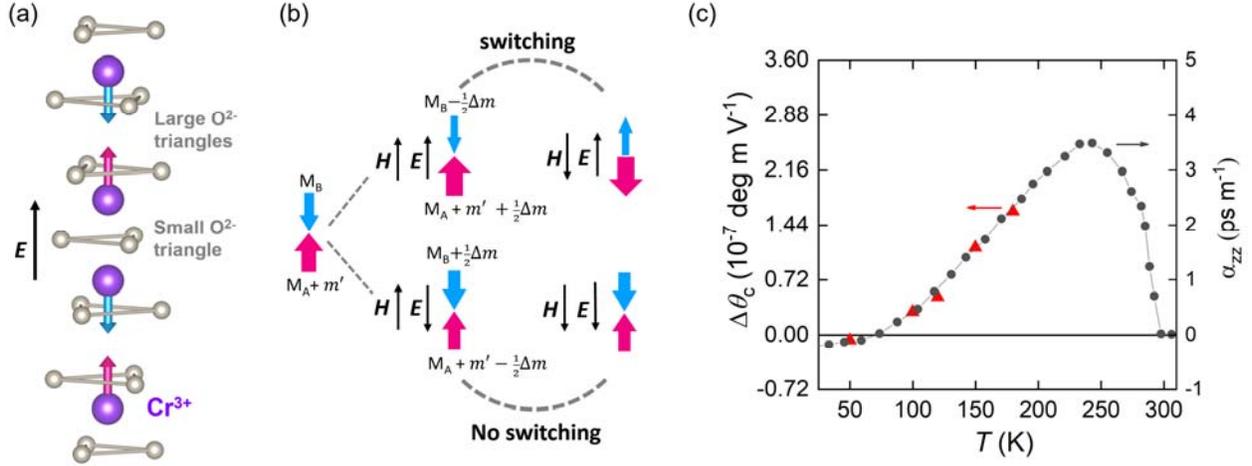

Figure 4. $Cr_2O_3$ lattice structure and ME manipulation of the magnetic sublattices. (a) Spin structure of $Cr^{3+}$ in a rhombohedral cell of $Cr_2O_3$. The two magnetic sublattices are shown as red (A) and blue (B) arrows, respectively. (b) Illustration of the voltage-induced switching of the magnetic sublattices via the ME effect. $M_A$ and $M_B$ are bulk magnetizations of the two sublattices. Application of an electric field along the c-axis increase/decrease their magnitudes by $\Delta m/2$, as shown by the width of the arrows. In addition, the Cr spins at the bottom $Cr_2O_3$/Pt interface, belonging to sublattice A carries an extra magnetization $m'$ as described in the text. With the application of a magnetic field, the magnetic sublattices reorient themselves to lower the total magnetic energy. In a positive electric field (upper part), the switching process is aided by the electric field induced magnetic moment. Conversely, a negative electric field (lower part) makes it harder to switch the magnetic sublattices. (c) Temperature dependence of the change of $\theta_c$ normalized by the electric field (left axis) and the magnitude of the ME coefficient $\alpha_{zz}$ (right axis). The data of $\alpha_{zz}$ is reproduced from ref. 42.

Figure 4(b) illustrates how the switching of the magnetic sublattices is induced by an electric field. To get an intuitive understanding of the different behaviors under a positive and negative bias voltages, we also need to consider that the $Cr_2O_3$ films in our SSE device is slightly 'ferrimagnetic'. Our magnetic-field cooling and angular dependence of SSE measurements indicate that the sublattice A in contact with the bottom Pt layer may carry an extra unbalanced magnetization $m'$ (Supplemental Materials). The $m'$ can come from extra Cr atoms formed at the bottom $Cr_2O_3$/Pt interface resulting from misfit dislocations during heteroepitaxial film growth [32]. More details on the imbalance between the sublattice A and B, including effect of magnetic anisotropies, are presented in Supplemental Note 1. Under a positive electric field, as shown in Fig. 4(b), the increased/decreased magnetization $\Delta m/2$ in the sublattice A/B from the ME effect produces a net magnetization $\Delta m$ that, adding to $m'$, allows the magnetic lattices to be switched more readily in response to the rotation of magnetic field to lower the magnetic energy. Under a negative bias voltage, however, the extra magnetization $m'$ opposes the ME-induced magnetization $\Delta m$. As shown in the lower part of Fig. 4(b), when $\Delta m$ is smaller than or comparable to $m'$, the switching process is hindered because the difference in magnetic energies between the two configurations of the magnetic sublattices is reduced. We have obtained a quantitative analysis of the electric-field dependent switching of magnetic sublattices in proximity to the spin-flop transition, by using a minimum model with two antiferromagnetically coupled magnetizations taking into account the change in magnetic energy density from the ME effect. The details are presented in Supplemental Note 2.

We performed measurements of the voltage control of the SSE at different temperatures. We found that the tunability of the switching angle $\theta_c$ has a strong temperature dependence, which is shown in Fig. 4(d). The change in $\theta_c$ normalized by the electric field is considerably suppressed at temperatures below 100 K, which further changes sign at 50 K. Remarkably, this behavior of the temperature dependence follows closely the temperature dependence [42] of the ME coefficient $\alpha_{zz}$ of $Cr_2O_3$, which is shown on the right axis in the same plot. These observations provide further evidence that voltage-controlled switching in the SSE comes from the ME coupling in $Cr_2O_3$.

In conclusion, we have investigated voltage control of the spin Seebeck effect using antiferromagnetic $Cr_2O_3$ thin films. The unique sensitivity of the SSE to the magnetic sublattice and the presence of a magnetoelectric effect in $Cr_2O_3$ enables direct control of the SSE signal by applying a bias voltage. We have presented a model based on the magnetoelectric effect, which explains the bias-voltage control of the SSE measurement. These experimental demonstrations open new avenues for controlling spin currents in antiferromagnets, creating exciting prospects for antiferromagnetic spintronics.

We thank Hilal Saglam and Yi Li for assistance in initial sample growth. All work at Argonne was supported by the US Department of Energy, Office of Science, Basic Energy Sciences, Materials Sciences and Engineering Division. The use of facilities at the Center for Nanoscale Materials, an Office of Science user facility, was supported by the US Department of Energy, Basic Energy Sciences under Contract No. DE-AC02-06CH11357. The work by S. Z. on the theoretical analysis of the voltage-dependent spin-flop field was supported by the College of Arts and Sciences, Case Western Reserve University.

Supplemental Materials for *"Voltage control of magnon spin currents in antiferromagnetic Cr$_2$O$_3$"*

Changjiang Liu, Yongming Luo, Deshun Hong, Steven S.-L. Zhang, Brandon Fisher, John E. Pearson, J. Samuel Jiang, Axel Hoffmann and Anand Bhattacharya

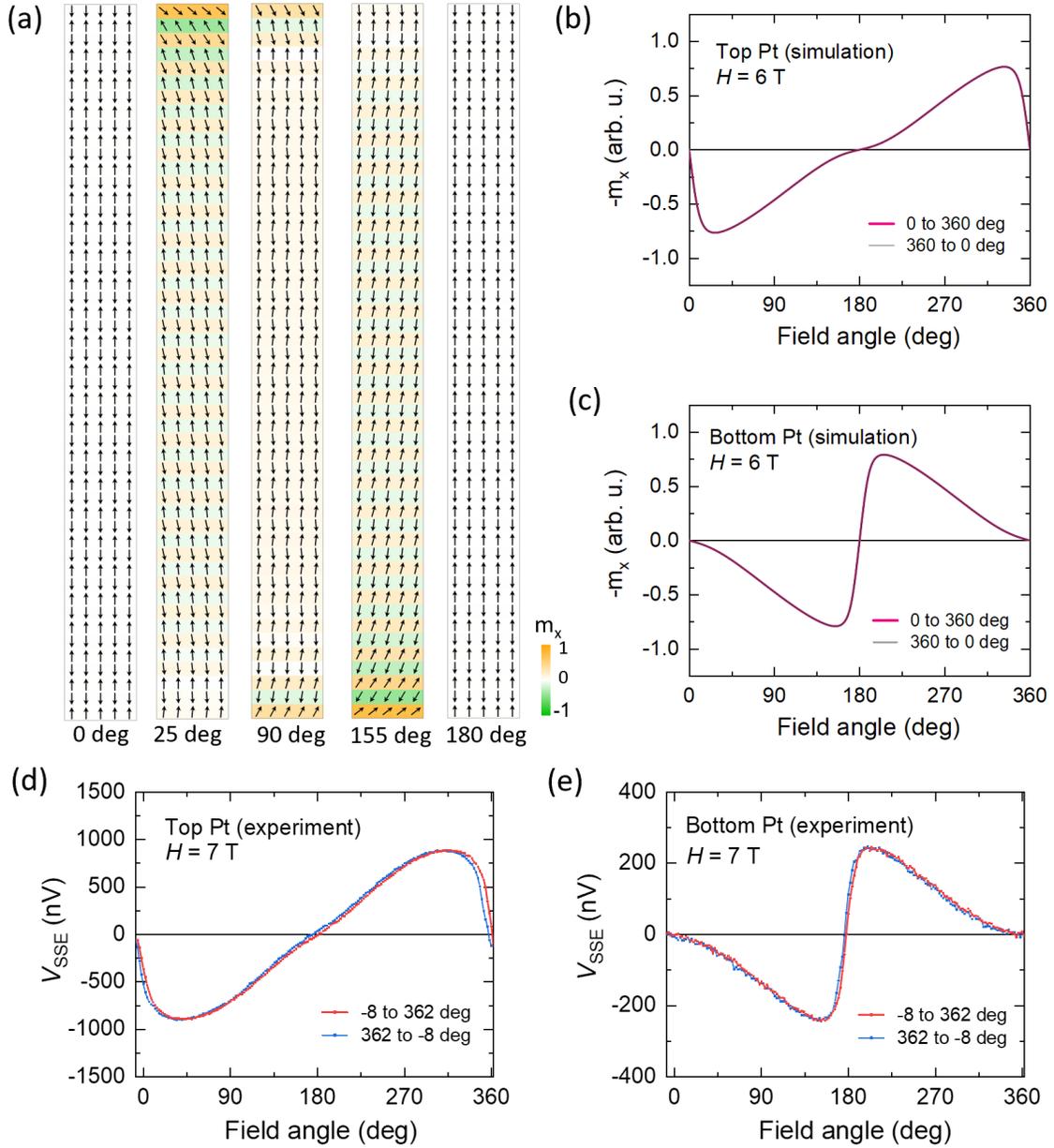

**Fig. S1**. (a) Micromagnetic simulation results of the spin orientations of the magnetic sublattices at a low magnetic field (6 T) for several different field angles. The simulation is performed using OOMMF by considering antiferromagnetic coupling ( $|A_{\mathrm{AF}}| = 4 \times 10^{-12}$ J/m) between adjacent layers, ferromagnetic coupling ($|A_{\mathrm{F}}| = 4 \times 10^{-12}$ J/m) within each layer, and a uniaxial magnetic anisotropy energy ($2 \times 10^4$ J/m$^3$) with easy axis along the z-direction. The cell size in the simulation is taken as 5× 5 × 5 nm. The region in the simulation has a dimension of 5 × 1 × 50 cells. The sublattice magnetization is taken as $M_A = (1 + 0.3\%) \times 22579$ A/m and $M_B = (1 − 0.3\%) \times 22579$ A/m, where the quantity 0.3% is an inserted parameter to match the experimental data. See also Supplemental Note 1. (b) and (c), the *x*-component of the magnetization in the top and bottom surface plotted as a function of the field angle, respectively. These reproduce the angular dependence of the SSE signal observed in the experiment, which are shown in (d) and (e) as measured in the top and bottom surfaces, respectively. Note that the different size of the raw SSE voltage in the top and bottom surfaces are due to the difference in the resistivity of the Pt detectors. The line shape in (d) and (e) corresponds to the initial orientation of the magnetic sublattice in contact with the bottom Pt being along positive *z*-direction as can be seen in (a) for field angle at 0 deg or 180 deg.

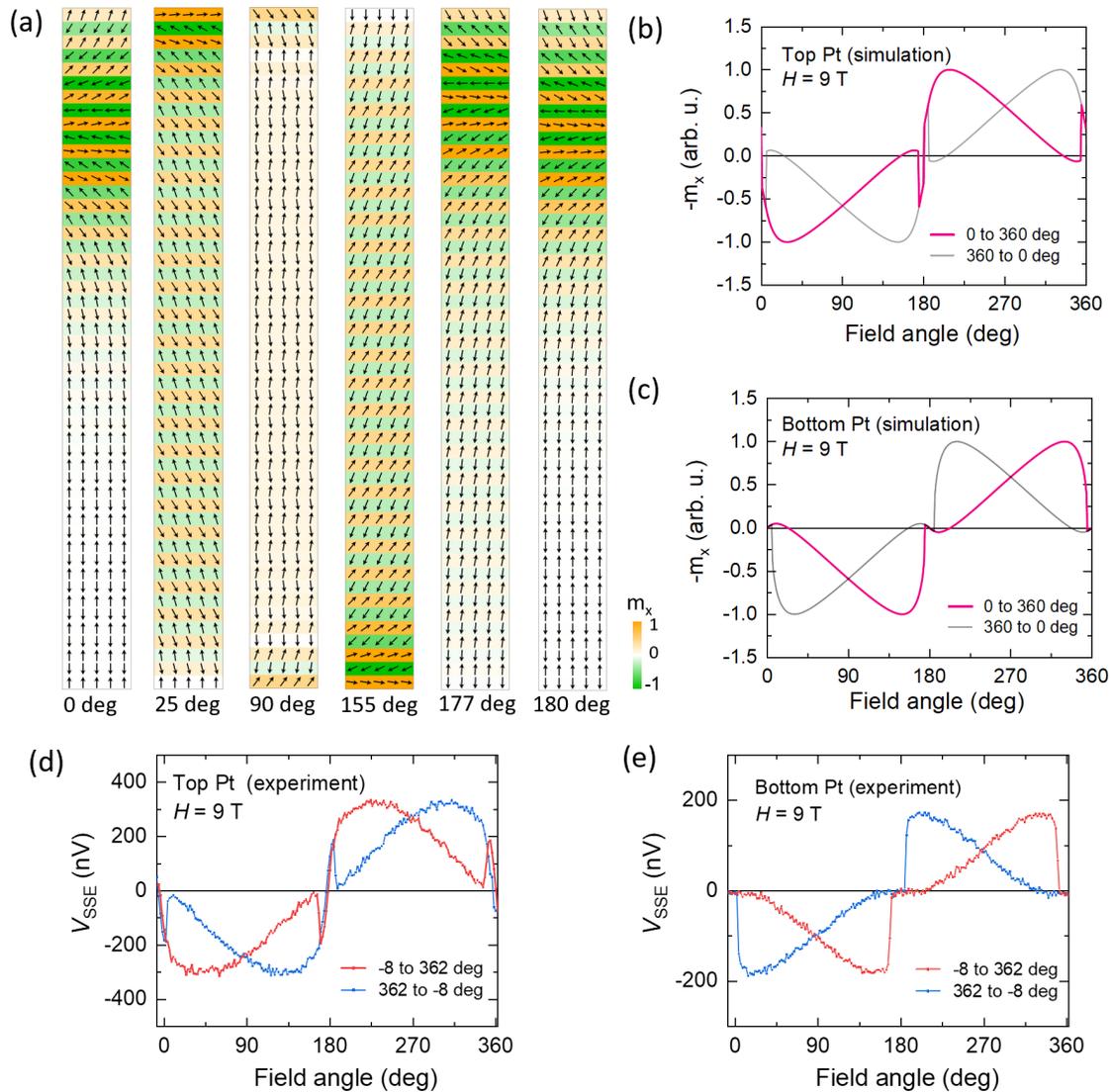

**Fig. S2**. (a) Simulation results of the spin orientations of the magnetic sublattices at a high field (9 T), above the surface spin-flop transition, for several different field angles. The spin-flopped states are seen in the upper part of the layers at magnetic field angles of 0 deg or 180 deg. (b) and (c),

the *x*-component of the magnetization in the top and bottom surface plotted as a function of the field angle, respectively. (d), (e) Angular dependence of the SSE signals measured in the top and bottom Pt layers, respectively, showing largely the same angular dependent line shapes as in the simulation results in (b) and (c).

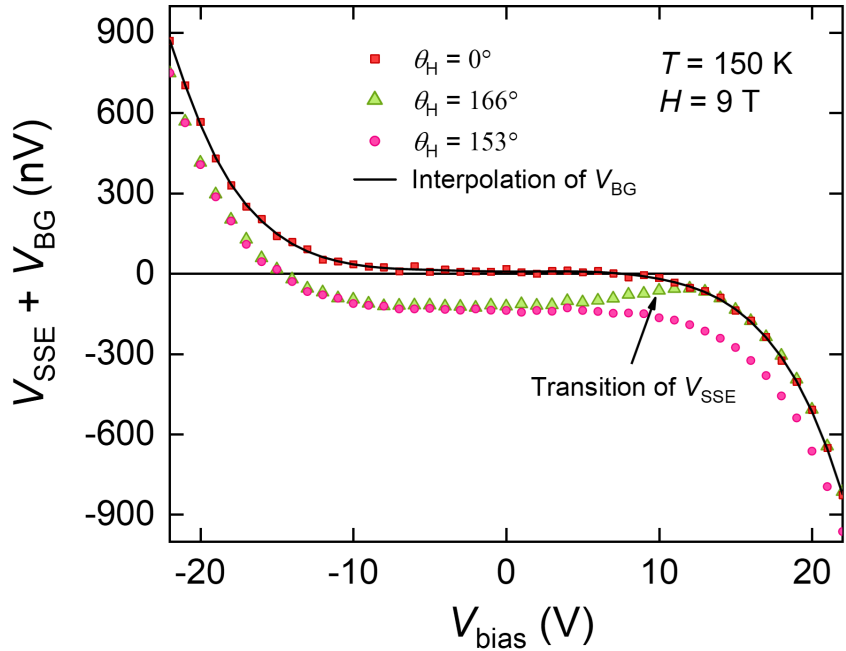

**Fig. S3**. Raw signal measured in the Pt layer as a function of bias voltage. The background voltage, $V_{BG}$, is obtained by measuring the voltage with field applied along the *c*-axis of $Cr_2O_3$ ($\theta_H = 0$ deg), where $V_{SSE} = 0$ V. The magnitude of $V_{BG}$ increases substantially when $V_{bias} > 15$ V due to the leakage current through the $Cr_2O_3$ film. The black solid curve is an interpolation to the bias voltage dependence of the $V_{BG}$. An example of the transition in SSE can be seen in the green data points, as indicated by the arrow in the plot. The $V_{SSE}$ shown in Fig. 3(b) of the main text are obtained by subtracting the bias voltage dependence of $V_{BG}$ from the total voltages measured in the Pt layer. The leakage current is significantly reduced at lower temperatures (< 120 K), and a subtraction of $V_{BG}$ is not necessary.

**Supplemental Note 1: Signs of 'ferrimagnetism' in $Cr_2O_3$ film**

In the SSE measurement, we found that the $Cr_2O_3$ film is slightly 'ferrimagnetic' even under zero bias voltage. This is identified by the preferred alignment of the magnetic sublattices under an application of only a magnetic field. As shown in Fig. S1, the orientations of the magnetic sublattices can be determined using the line shapes of the angular dependence of the SSE at low field. Our measurements show that the sublattice in direct contact with the bottom Pt is always aligned to be parallel with the magnetic field. This alignment is obtained when only a magnetic field is applied along the *c*-axis of $Cr_2O_3$ film while the sample is cooled down through the Néel temperature; or when a large magnetic field (9 T, above the spin-flop transition field) is applied along *c*-axis and subsequently ramped down in the temperature range where the SSE measurement is performed.

The slight 'ferrimagnetism' in $Cr_2O_3$ film may be due to a boundary magnetization that exists at the bottom interface ($Cr_2O_3$/Pt/$Al_2O_3$). Similar observation has been reported previously (ref. 32 of the main text), in which the boundary magnetization is attributed to extra Cr atoms at the interface from misfit dislocations during the epitaxial growth. We note that in our samples the bottom $Cr_2O_3$/Pt/$Al_2O_3$ interface is grown at high temperature ($Cr_2O_3$ growth temperature), while the top $Cr_2O_3$/Pt interface is formed at room temperature.

Additionally, based on our simulations as shown in Fig. S2, the hysteretic angular dependence of the $V_{SSE}$ can only be obtained when there is a slight imbalance between the two magnetic sublattices. The imbalance can be: a slightly higher magnetization (be in the bulk or at the boundary) in the sublattice in contact with the bottom Pt, a higher magnetic anisotropy energy near the bottom $Cr_2O_3$/Pt interface and/or a reduced magnetic anisotropy in the top-surface layers. We note that the difference in magnetic anisotropy along the thickness of $Cr_2O_3$ also help set a preferred orientation of the magnetic sublattices under an applied magnetic field, as found in our simulation. By comparing the line shapes obtained from the simulation (Fig. S2) for both the bottom and top surfaces under these different conditions, we found that a slight imbalance in the bulk magnetic moments (~ 0.3%) produces the angular dependent curve mostly similar to the experimental results. The imbalance in bulk magnetic moments could arise from the non-zero longitudinal magnetic susceptibility of $Cr_2O_3$, which is not considered in our simulation.

**Supplemental Note 2: Electric-field dependent spin-flop in $Cr_2O_3$**

The switching of the SSE in the angular dependence measurement is accompanied with a "spin flop" in the magnetic sublattices, which can be seen in Fig. S2(a) with the magnetic field angle increases from 155 deg to 177 deg. The orientations of the magnetic sublattices in the upper part of the sample (also shown in Fig. S4) changes from being largely aligned with the *c*-axis to perpendicular to it (spin-flopped state). In our measurement, we found that applying an electric field can either promote or inhibit this transition as shown in Figs. 2 and 3 of the main text.

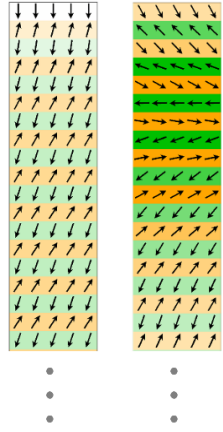

**Fig. S4.** Orientations of the magnetic sublattices before (left) and after (right) the switching in the SSE.

In the following, we use a macrospin approximation to show the dependence of the spin-flop field on the external electric field due to the magnetoelectric (ME) effect and provide an order-of-magnitude estimation of the correction to the spin-flop field. The extra ferrimagnetic magnetization $m'$ described in the main text is not considered in this minimum model. To derive the spin-flop field, we consider both the applied electric field $\mathbf{E}$ and magnetic field $\mathbf{H}_a$ are collinear with the $c$-axis of a $Cr_2O_3$ lattice. Due to the linear ME effect, the electric field gives rise to a net magnetization, i.e.,

$$\mu_0 \delta M_z = \alpha_{zz} E_z,\ \delta M_x = \delta M_y = 0, \tag{S1}$$

where we have assumed the $c$-axis to be parallel to the $z$-axis in the present geometry (note that with both $\mathbf{E}$ and $\mathbf{H}_a$ fields being collinear with the $c$-axis, the ME free energy density term is reduced to $-\alpha_{zz} H_{a,z} E_z$ and hence only the $z$-component of the induced magnetization is nonzero). The sign of $\alpha_{zz}$ depends on the spin configuration determined in field cooling, as explained in the main text. We also assume the magnetization magnitudes of the two sublattice are increased and decreased by the same amount respectively, i.e.,

$$\mathbf{M}_A = (M_s + \tfrac{1}{2}\delta \widetilde{M}_s)\widehat{\mathbf{M}}_A \text{ and } \mathbf{M}_B = (M_s - \tfrac{1}{2}\delta \widetilde{M}_s)\widehat{\mathbf{M}}_B \tag{S2}$$

with $M_s$ ($\equiv |\mathbf{M}_A^{(0)}| = |\mathbf{M}_B^{(0)}|$) being the magnitude of sublattice magnetization in the absence of the electric field, and $\hat{\mathbf{M}}_A$ and $\hat{\mathbf{M}}_B$ being the unit vectors denoting the directions of the sublattice magnetizations. By symmetry,

$$M_{A,z} + M_{B,z} = M_s(\hat{M}_{A,z} + \hat{M}_{B,z}) + \frac{1}{2}\delta\tilde{M}_s(\hat{M}_{A,z} - \hat{M}_{B,z}) = \delta M_z. \quad (S3)$$

Note that from Eq. (S3) we can see that while $\delta M_z$ is a constant for fixed $\mathbf{E}$ and $\mathbf{H}_a$ fields, $\delta\tilde{M}_s$ in general depends on the magnetic configuration of the system (it becomes the same as $\delta M_z$ only in the antiferromagnetic state when both sublattice magnetizations are collinear with the c-axis).

The magnetic energy density can be written as

$$\epsilon_M = -H_a(M_{A,z} + M_{B,z}) - \frac{H_K}{2M_s}(M_{A,z}^2 + M_{B,z}^2) + \frac{H_J}{M_s}(\mathbf{M}_A \cdot \mathbf{M}_B), \quad (S4)$$

Where $H_K$ and $H_J$ are the effective anisotropy and exchange fields, which are both positive. Without loss of generality, we assume $\mathbf{M}_A$, $\mathbf{M}_B$, and $\mathbf{H}_a$ lie in the same plane so that the directions of $\mathbf{M}_A$ and $\mathbf{M}_B$ can be described by their polar angles. It follows that the magnetic energy density can be expressed as

$$\epsilon_M(\theta_A, \theta_B)/M_s = -H_a\left[\cos\theta_A + \cos\theta_B + \frac{1}{2}\delta\tilde{m}_s(\cos\theta_A - \cos\theta_B)\right]$$

$$-\frac{1}{2}H_K\left[\left(1 + \frac{1}{2}\delta\tilde{m}_s\right)^2 \cos^2\theta_A + \left(1 - \frac{1}{2}\delta\tilde{m}_s\right)^2 \cos^2\theta_B\right]$$

$$+H_J\cos(\theta_A + \theta_B)\left[1 - \frac{1}{4}(\delta\tilde{m}_s)^2\right], \quad (S5)$$

where $\delta\tilde{m}_s (\equiv \delta\tilde{M}_s/M_s)$ is a dimensionless quantity characterizing the ratio of the magnitude of the induced moment to $M_s$. Keeping terms up to the first order of the electric field, we get

$$\epsilon_M(\theta_A, \theta_B)/M_s \stackrel{O(\delta\tilde{m}_s)}{\simeq} -H_a(\cos\theta_A + \cos\theta_B) - \frac{1}{2}H_K(\cos^2\theta_A + \cos^2\theta_B)$$

$$+H_J\cos(\theta_A + \theta_B) - \frac{1}{2}\delta\tilde{m}_s[H_a(\cos\theta_A - \cos\theta_B) + H_K(\cos^2\theta_A - \cos^2\theta_B)]. \quad (S6)$$

Stable magnetic configurations $(\theta_A^m, \theta_B^m)$ corresponds to magnetic energy minima, which must satisfy

$$\frac{\partial}{\partial \theta_A} \epsilon_M(\theta_A^m, \theta_B^m) = \frac{\partial}{\partial \theta_B} \epsilon_M(\theta_A^m, \theta_B^m) = 0 \tag{S7a}$$

and

$$\mathcal{A}(\theta_A^m, \theta_B^m)\mathcal{C}(\theta_A^m, \theta_B^m) - [\mathcal{B}(\theta_A^m, \theta_B^m)]^2 > 0 \text{ with } \mathcal{A} > 0 \text{ (or } \mathcal{C} > 0\text{)}, \tag{S7b}$$

where $\mathcal{A} \equiv \frac{\partial^2 \epsilon_M}{\partial \theta_A^2}$, $\mathcal{C} \equiv \frac{\partial^2 \epsilon_M}{\partial \theta_B^2}$, and $\mathcal{B} \equiv \frac{\partial^2 \epsilon_M}{\partial \theta_A \partial \theta_B}$.

Placing Eq. (S6) in (S7), the stationary points can be determined by the following equations

$$H_a \sin \theta_A + \frac{1}{2} H_K \sin 2\theta_A - H_J \sin(\theta_A + \theta_B) + \frac{1}{4} \delta\widetilde{m}_s (2H_a \sin \theta_A + H_K \sin 2\theta_A) = 0, \tag{S8a}$$

$$H_a \sin \theta_B + \frac{1}{2} H_K \sin 2\theta_B - H_J \sin(\theta_A + \theta_B) - \frac{1}{4} \delta\widetilde{m}_s (2H_a \sin \theta_B + H_K \sin 2\theta_B) = 0. \tag{S8b}$$

At the zeroth order of $\delta\widetilde{m}_s$, we find the known spin-flop angle (same for the magnetizations of both sublattices)

$$\theta_A^{(0)} = \theta_B^{(0)} = \theta_{sf} = \cos^{-1}\left(\frac{H_a}{2H_J - H_K}\right). \tag{S9}$$

To solve for the first-order terms, we let

$$\theta_A \simeq \theta_A^{(0)} + \delta\theta, \text{ and } \theta_B \simeq \theta_B^{(0)} - \delta\theta, \tag{S10}$$

so that $\sin(\theta + \delta\theta) \simeq \sin\theta + \delta\theta \cos\theta$, and $\cos(\theta + \delta\theta) \simeq \cos\theta - \delta\theta \sin\theta$. Note that since the magnitude of the sublattice magnetization are now different, the changes in the tilting angles of the magnetizations in the spin-flopped state should also be opposite in order to suppress a net transverse magnetization, as imposed by the symmetry (when both magnetic and electric fields are collinear with the c-axis, only the $\alpha_{zz}$ component enters the free energy and hence no net transverse magnetization is allowed). Placing Eq. (S10) in (S8), we get

$$H_a \delta\theta \cos\theta_A^{(0)} + H_K \delta\theta \cos 2\theta_A^{(0)} + \frac{1}{2}\delta\widetilde{m}_s(H_a \sin\theta_A^{(0)} + H_K \sin\theta_A^{(0)} \cos\theta_A^{(0)}) = 0 \tag{S11a}$$

$$-H_a \delta\theta \cos\theta_B^{(0)} - H_K \delta\theta \cos 2\theta_B^{(0)} - \frac{1}{2}\delta\widetilde{m}_s(H_a \sin\theta_B^{(0)} + H_K \sin\theta_B^{(0)} \cos\theta_B^{(0)}) = 0, \tag{S11b}$$

which can be solved exactly and give

$$\delta\theta = \frac{\delta\widetilde{m}_s H_J H_a \sqrt{(2H_J - H_K)^2 - H_a^2}}{H_a^2 (2H_J + H_K) - H_K (2H_J - H_K)^2}. \tag{S12}$$

Keeping terms up to the first-order in $\frac{H_a}{H_J}$ and $\frac{H_K}{H_J}$, we arrive at

$$\delta\theta \simeq -\left(\frac{H_a}{2H_K}\right) \delta\widetilde{m}_s. \tag{S13}$$

Note that this result indicates that the uncompensated magnetizations of the two sublattices acquire slightly different tilting angles about the $c$-axis.

The magnetic system has two energy minima:

1) $\epsilon_M^{AF}/M_s \stackrel{O(\delta\widetilde{m}_s)}{\simeq} -\delta\widetilde{m}_s^{AF} H_a - H_K - H_J$ (S14)

(if $\delta\widetilde{m}_s H_a > 0\ (< 0)$, then energy minimum is at $(0, \pi)$ $[(\pi, 0)]$), which corresponds to antiparallel alignment of $\mathbf{M}_A$ and $\mathbf{M}_B$, and

2) $\epsilon_M^{SF}(\theta_{sf} + \delta\theta, \theta_{sf} - \delta\theta)/M_s \stackrel{O(\delta\widetilde{m}_s)}{\simeq} -H_J - \frac{H_a^2}{2H_J - H_K},$ (S15)

which corresponds to a spin-flop state with unequal tilting angles for the sublattice magnetizations, i.e.,

$$\theta_{sf}^A \simeq \theta_{sf} + \delta\theta \text{ and } \theta_{sf}^B \simeq \theta_{sf} - \delta\theta \tag{S16}$$

with the zeroth-order spin-flop angle, $\theta_{sf}$, and its first-order correction, $\delta\theta$, given by Eq. (S12) or (S13), respectively. Note that $\epsilon_M^{SF}$ remains the same as that in the absence of electric field [as shown by Eq. (S15)], as the correction to the energy density due to the ME effect is of second-order in $\delta\widetilde{m}_s$.

Letting $\epsilon_M^{AF}(0, \pi) = \epsilon_M^{SF}(\theta_{sf} + \delta\theta_{ME}, \theta_{sf} - \delta\theta_{ME})$, we obtain the critical field for the spin-flop transition as

$$H_{sf} \stackrel{O(\delta\widetilde{m}_s)}{\simeq} \sqrt{H_K(2H_J - H_K)} + \frac{1}{2}\delta\widetilde{m}_s^{AF}(2H_J - H_K), \tag{S14}$$

Placing Eq. (S1) in the above equation (recall that in the AF state we have $\delta \widetilde{M}_s = \delta M_s$ and thus $\delta \widetilde{m}_s^{AF} = \alpha_{zz}^{AF} E_z / \mu_0 M_s$), we arrive at our final expression for the spin-flop field in the presence of an external electric field, i.e.,

$$H_{sf} \stackrel{O(\delta \widetilde{m}_s)}{\simeq} \sqrt{H_K(2H_J - H_K)} + \frac{\alpha_{zz}^{AF} E_z}{2\mu_0 M_s}(2H_J - H_K). \tag{S15}$$

Note that, up to the first-order in the electric field $E_z$, the correction to the spin-flop field is proportional to the ME coefficient for the AF state and is independent of that for the spin-flop state (as the latter is of second-order in $E_z$). Using parameters for $Cr_2O_3$ (refs. 42 and 44 of the main text) : $\alpha_{zz}^{AF}(150\ K) = 2$ ps/m, $M_s = 2.86 \times 10^5$ A/m, $V_{bias} = 24\ V$, $d = 250$ nm, $E_z \left(= \frac{V_{bias}}{d}\right) = 8.3 \times 10^7$ V/m, $H_J = 245$ T, and $H_K = 0.07\ T$, we find

$$\Delta H_{sf} = \frac{\alpha_{zz}^{AF} E_z}{2\mu_0 M_s}(2H_J - H_K) \sim 0.13\ \text{T}.$$

In our measurement, as shown in Fig. 3 of the main text, the critical angle $\theta_c$ decreases from 171 deg to 162 deg when $V_{bias}$ increases from 0 V to 24 V. This corresponds to a decrease in the spin-flop field

$$\Delta H_{sf}^{exp} = 9[|\cos \theta_c(0\ V)| - |\cos \theta_c(24\ V)|] \sim 0.33\ \text{T},$$

which is of the same order-of-magnitude as the theoretical estimation.

**Author contributions**:

C.J.L. performed the SSE measurements and established the voltage-control of the SSE in $Cr_2O_3$. C.J.L. and Y.M.L grew and fabricated the devices. D.S.H. performed x-ray characterizations of the sample. C.J.L. performed the micromagnetic simulations presented in this paper. S.Z. provided theoretical analysis of the electric-field dependent spin-flop transition. B.F., J.E.P and J.S.J provided assistance during the experiment. All authors were involved in the data analysis, discussion, and manuscript preparation. The project was supervised by A.B. and A.H..